\documentclass{article}

\usepackage[left=2cm,top=2.5cm,right=2cm,bottom=3cm,nohead]{geometry}
\usepackage{amsmath,amsfonts,amsthm,amssymb}
\usepackage{epsfig}
\usepackage{color,verbatim,epic, eepic}

\newcommand{\Ht}{\mathbb{H}^3}
\newcommand{\dif}[2]{\frac{d #1}{d #2}}   
\newcommand{\pd}[2]{\frac{\partial #1}{\partial #2}}

\def\ben{\begin{equation}}
\def\een{\end{equation}}
\def\bea{\begin{eqnarray}}
\def\eea{\end{eqnarray}}
\def\gp{\wp}

\numberwithin{equation}{section}
\begin{document}

\hfuzz=100pt
\title{Light-bending in Schwarzschild-de-Sitter: 
projective geometry of the optical metric}
\author{G W Gibbons$^{A,}$\footnote{g.w.gibbons@damtp.cam.ac.uk}\ , C M Warnick$^{A,}$\footnote{c.m.warnick@damtp.cam.ac.uk}\ \ \& M C Werner$^{A, B,}$\footnote{mcw36@cam.ac.uk}}
\date{}

\maketitle

\vspace{-.6cm}

\centerline{ ${}^A$ DAMTP, University of Cambridge, Wilberforce Road, Cambridge CB3 0WA, UK}

\centerline{ ${}^B$ IoA, University of Cambridge, Madingley Road, Cambridge CB3 0HA, UK}

\vspace{-6.5cm}

\begin{flushright}
DAMTP-2008-74
\end{flushright}

\vspace{6cm}

\begin{abstract} 
We interpret the well known fact that the equations for light rays in the Kottler or Schwarzschild-de Sitter metric are independent of the cosmological constant in terms of the projective equivalence of the optical metric for any value of $\Lambda$. We explain why this does not imply that lensing phenomena are independent of $\Lambda$. Motivated by this example, we find a large collection of one-parameter families of projectively equivalent metrics including both the Kottler optical geometry and the constant curvature metrics as special cases. Using standard constructions for geodesically equivalent metrics we find classical and quantum conserved quantities and relate these to known quantities.
\end{abstract}

\vspace{1cm}

\section{Introduction}

Some time ago, Islam \cite{Islam} observed that the differential equation
\ben
\frac{d ^2 u}{d \phi ^2 }+u = 3 m u^2\,  ,\label{diff} 
\een
where $u=1/r$, and  $r$ is the usual Schwarzschild radial
coordinate, which   
governs  light rays moving in a Schwarschild-de-Sitter, or 
 Kottler \cite{Kottler},
metric 
depends  only on the mass $m$ and not  the cosmological constant $\Lambda$\footnote{Throughout we take units such that $c=G=1$}.

One might therefore  be led into supposing  that
the light deflection and time-delay formulae, which relate
observable quantities  
would be the same as in the Schwarzschild metric, or because
our present universe appears to be dominated by a cosmological term,
that the usual formulae for  gravitational lensing in Friedmann-Lema\^itre
universes are not applicable if the cosmological constant 
is non-zero. More excitingly, one might suppose
that gravitational lensing observations could allow
a new and  independent measurement of the cosmological constant.  

There has been a considerable amount of  discussion
of these  possibilities recently and the consensus
is emerging that as far as the  observational situation
is concerned, the existing theory of gravitational lensing
is perfectly adequate (see \cite{Sereno:2007rm, Park} and references therein). There remain however questions
about precisely what equation (\ref{diff}) is telling one
and why it is misleading.

Roughly speaking, the origins of the   confusion are two-fold
\begin{itemize}

\item   Equation (\ref{diff}) merely governs the unparameterized  projection
of null rays onto the  spatial sections of the Kottler metric
and not more detailed geometrical features such as  deflection  angles
and  differences in duration of different paths
joining a source to an observer.

\item In the cosmological setting, the  source and observer
both participate in the Hubble flow and hence  are 
not at rest with respect to the static Kottler coordinates,
but are in relative motion, and this must also  be taken into account.

\end{itemize}

It is well known that according to Fermat's principle (e.g., \cite{Synge}),
the projection  of null geodesics
onto a surface of constant time are geodesics of 
a conformally rescaled spatial  metric called 
the optical metric. One of the principle  purposes of the present paper
is to explain that Islam's observation is essentially
that the projective properties of these geodesics are independent
of the cosmological constant, but not their  lengths nor their 
conformal properties such as angles. In fact the optical metrics of the Kottler spacetimes provide an hitherto unrecognized example of a one parameter
family of projectively equivalent metrics, which may be of interest to 
geometers in its own right.

The second aspect of the problem relates to the full four-dimensional
spacetime geometry. As was first realized by McVittie \cite{McVittie},
the Kottler metric may be cast in a form which clearly exhibits
it as the gravitational field  of a point mass at respect with respect 
to spatially flat Friedmann-Lema\^itre expanding universe with de-Sitter 
type exponentially expanding  scale factor     
\ben
a(t) = \exp( H t) \,,\qquad H= \sqrt{ \frac{3 }{ \Lambda}}\,.
\een
 
Since, to lowest order in the mass $M$, this coincides precisely
with the standard form of the approximate metric used in all studies
of   gravitational lensing, it is clear that no 
new or unaccounted effects can be obtained working to linear order in $M$
using the Kottler form which cannot also be obtained using the standard 
linearized McVittie form. This point has been made
forcibly recently by Park \cite{Park} who shows explicitly   
that the light deflection formulae, if derived correctly,
are are identical in both formalisms. 

In this paper we shall first establish the lensing problem in terms of sources and observers fixed in coordinates comoving with the Hubble flow and demonstrate that one may instead consider a moving source and observer in the Kottler metric. By passing to the optical metric we will derive the standard equations governing light rays.  We will then discuss the geometrical notion of projective equivalence and show that this explains the peculiar property of equation (\ref{diff}). Led by the Kottler metrics, we find a large class of families of geodesically equivalent metrics and using results from the literature exhibit conserved quantities of their geodesic flow.

\section{McVittie's formalism and the Kottler metric}

In order to calculate the effects of gravitational lensing
we need the  metric due to a system of co-moving mass points in
a background Friedman-Lema\^itre universe. This was  given to lowest 
order  by McVittie in 1964
\cite{McVittie}. McVittie's approximate metric is
\ben
ds^2 = -\left( 1 + {2 U} \over {a^2} \right)dt^2 + a^2
\left( 1 - {2U} \over {a^2} \right)
\left\{ { d {\bf x } ^2 \over { ( 1+ k{ {\bf x }^2 \over 4 l^2} )}} \right\},
\een
where $k=-1,0,1$, depending on whether the metric inside the brace is
hyperbolic space, Euclidean space or spherical space and $L$ is the radius of curvature of these spaces. The  Newtonian potential $U({\bf x})$ satisfies
\ben
\nabla^2_k U=0,
\een
where $\nabla ^2_k$ is the Laplacian with respect to the metric
in the brace.

A typical physical question concerns the light deflection caused by a lensing galaxy in a space with a positive cosmological constant. In order to study this, one may make use of McVittie's version of the Kottler or Schwarzschild-de Sitter metric. In McVittie's coordinates the metric takes the form:
\ben
ds^2 = - \left(\frac{1-\frac{m}{2 a(t) x}}{1+\frac{m}{2 a(t) x}} \right ) dt^2 + a(t)^2 \left(1+\frac{m}{2 a(t) x} \right)^4 \left\{ dx^2 + x^2 \left(d\theta^2 + \sin^2 \theta d \phi^2 \right)\right \}
\een
where $a(t) = \exp(t/l)$. This represents an exact solution of Einstein's equations with a cosmological constant $\Lambda = 3/l^2$. This form is of interest because for $x$ large it approaches the standard FRLW form of the de Sitter metric familiar to cosmologists. Sources and observers are considered to be fixed in the comoving $x, \theta, \phi$ coordinates. We may make a coordinate transformation to the static Kottler form of the metric as follows:
\ben
T =  (t + \tau(r)), \qquad r = e^{t/l} x + m + \frac{m^2}{4 e^{t/l}x} \label{Teqn}
\een
where the function $\tau$ is defined by:
\ben
\dif{\tau}{r} = - \frac{r^2}{l (r-m)(1-\frac{m}{r}-\frac{r^2}{l^2})} \label{taueqn}
\een
This gives the metric form:
\ben
ds^2 = -\left (1-\frac{2m}{r} - \frac{r^2}{l^2} \right) dT^2 + \frac{dr^2}{\left (1-\frac{2m}{r} - \frac{r^2}{l^2} \right)} + r^2\left(d\theta^2 + \sin^2 \theta d \phi^2 \right)
\een
The advantage of these coordinates is that they are static, making the problem of finding light rays simpler. The disadvantage is that observers `at rest' in the comoving coordinates are in motion with respect to the Kottler coordinates. One may readily translate between the two pictures using equations (\ref{Teqn}, \ref{taueqn}).

It is convenient to pass to the optical metric in order to find the light rays. This is given by:
\ben
ds^2_{\mathrm{opt}.} = \frac{dr^2}{\left (1-\frac{2m}{r} - \frac{r^2}{l^2} \right)^2} + \frac{r^2}{\left (1-\frac{2m}{r} - \frac{r^2}{l^2} \right)}\left(d\theta^2 + \sin^2 \theta d \phi^2 \right)
\een
The geodesics of the optical metric are the projections of the light rays of the full spacetime onto a constant $T$ hypersurface, and length in the optical metric corresponds to the coordinate time $T$ taken by a light ray to traverse the line segment.

\subsection{Light Ray Equations}
From the Kottler optical metric we deduce the equations governing the light rays, which we make take without loss of generality to be in the equatorial plane, are:
\begin{equation}
\dif{\phi}{T} = \frac{\Delta_r h}{r^4}, \qquad \frac{r^4}{\Delta_r^2}\left( \dif{r}{T}\right)^2 + \frac{\Delta_r}{r^4} h^2 = 1
\end{equation}
Where
\ben
\Delta_r = r^2 - 2 m r - \frac{r^4}{l^2}
\een
Eliminating the time coordinate $T$ we arrive at an equation for the projection of the light rays into the $(r,\phi)$ plane:
\ben
\left(\dif{r}{\phi} \right)^2 + \Delta_r = \frac{r^4}{h^2}. \label{projgeod}
\een
We notice that by defining a modified angular momentum $k = hl/\sqrt{h^2+l^2}$ we may write this as:
\ben
\left(\dif{r}{\phi} \right)^2 + r^2 - 2 m r = \frac{r^4}{k^2}
\een
so that there is an equivalence between the light rays for different values of the cosmological constant, provided they give rise to the same value of $k$. We make the standard substitution $u=1/r$ to get the equations we will require:
\ben
\left(\dif{u}{\phi} \right)^2 = 2 m u^3 - u^2 + \frac{1}{k^2}, \qquad \dif{\phi}{T} = \left(u^2 - 2 m u^3 - \frac{1}{l^2} \right) h \label{geod}
\een
so that although the rays are the same, the length in the optical
metric (i.e.\ the time to traverse the ray) does depend upon the value
of the cosmological constant. There are other effects due to the fact
that the source and observer may not be taken to be at rest in the
static Kottler coordinates. We may in fact find solutions to the equation (\ref{projgeod}) using
elliptic functions:
\ben
{ m \over 2r(\phi) }- { 1 \over 12} = \gp(\phi + {\rm constant}),\,
\een 
where Weirstrass's  function $\gp $ satisfies
\ben
{\gp ^\prime } ^2 = 4 \gp ^3 -g_2 \gp -g_3\,,
\een
with
\ben
g_2= { 1 \over 12}\,,\qquad g_3= 
{ 1 \over 216} -{m^2 \over 4k^2 } \,.
\een

One may check that
\ben
u={ 1 \over 3m} - { 1 \over m} { 1 \over \cosh \phi +1} \label{cosh}
\een
and
\ben
u= { 1 \over m} { 1 \over \cos \phi +1}
\een
are exact solutions. The former(\ref{cosh}) starts at infinity
at $\phi= \cosh^{-1} (2)$  and moves inwards, spirally around the
circular
orbit at $r=3m$.   

In fact (\ref{cosh}) is not a Weierstrass
function
but if one adds $i{\pi \over 2}$ to the argument $\phi$ one gets
\ben
u= { 1 \over 3m}  + { 1 \over m} { 1 \over \cosh \phi -1} \,,
\label{sinh}\een 
which is a Weierstrass function.
This orbit starts from $r=0$  at $\phi=0$ and
moves outwards ultimately  endlessly approaching the circle
at $r=3m$. In practice, it is not convenient to work with these exact solutions
and an approximation is usually made.

\section{Projective Equivalence}

As Islam observed, the problem of finding the paths of null
geodesics in the Kottler spacetime is equivalent to the same problem
in the Schwarzschild spacetime. The reason for this is the
\emph{projective equivalence} of the optical metrics, which we define
as follows.

Given manifolds $\mathcal{M}^n$, $\mathcal{\bar{M}}^n$, two metrics
$g$, $\bar{g}$ are said to be projectively equivalent in
neighbourhoods $U$, $\bar{U}$ if there exist charts $\phi:U\to V$,
$\bar{\phi}:\bar{U}\to V$ with $V\subset \mathbb{R}^n$ such that the
geodesics of both metrics coincide as unparameterised curves on
$V$. We will refer loosely to geodesics as curves without any
prejudice as to their parameterization. For example, we may take
$\mathcal{M}=\mathbb{R}^3$, $g=\delta$ the Euclidean metric and
$\bar{g}$ to be the Beltrami metric for one half of the sphere:
\begin{equation}
\bar{g} = \frac{dr^2}{(1+\frac{r^2}{R^2})^2}+  \frac{r^2}{1+\frac{r^2}{R^2}}\left( d\theta^2 + \sin^2\theta d\phi^2\right)
\end{equation}
One may verify that the geodesics of this metric are in fact straight lines. Of course if we parameterize these curves by arc-length then they no longer agree with the Euclidean geodesics parameterized by arc-length since, for example, the length of a geodesic with respect to $\bar{g}$ may not exceed $\pi R$. Thus $g$ and $\bar{g}$ are projectively equivalent.

We may also consider $R \to i R$ which takes the metric $\bar{g}$ to the Beltrami metric for Hyperbolic $3$-space, $\Ht$. In this case we should be careful as these metrics are only defined inside the ball $r<R$, however the geodesics are still straight lines, so with some care regarding the domain of definition, these metrics are projectively equivalent to one another and to the flat metric.

In general, finding the coordinate charts $\phi$, $\bar{\phi}$ is fundamental to constructing a projective equivalence, since two different coordinate representations of the same metric may look quite different. We will for the moment take the question of the existence of charts as solved and restrict ourselves to the simpler question of whether two metrics defined on some subset of $\mathbb{R}^n$ have the same geodesics. In this case, an answer has been given by Eisenhart \cite[p131]{Eisenhart}. He shows by considering the geodesic equations for both metrics that a necessary condition for the geodesics to coincide is that
\begin{equation}
W^l{}_{ijk} = \bar{W}^l{}_{ijk}
\end{equation}
where the projective curvature tensor $W$ of $g$ is defined by:
\begin{equation}
W^l{}_{ijk} = R^l{}_{ijk} + \frac{1}{n-1}\left ( \delta^l{}_k R_{ij} - \delta^l{}_j R_{ik}\right).
\end{equation}
The index structure is important here, as we have two means of raising and lowering indices, with $g$ and $\bar{g}$, so should be careful\footnote{We take the curvature conventions $\left[ \nabla_i, \nabla_j \right ] X^k = R^k{}_{lij}X^l$ and $R_{ij} = R^k{}_{ikj}$, which differs from that of Eisenhart, who accordingly has a formula with $R_{ij} \to - R_{ij}$}. If $W$ vanishes for a spacetime, then by contracting indices one discovers that the space has constant curvature, so the only projectively flat spaces are those with constant curvature. In fact, as we saw above the spaces of constant curvature are indeed projectively equivalent.

\section{Projective Tensor for a warped product metric}

In addition to the projectively equivalent constant curvature metrics, we know that the Kottler metrics are a family of projectively equivalent metrics for all values of the cosmological constant. Motivated by this observation, we will consider a warped metric on the $n+1$ dimensional manifold $\mathbb{R} \times \mathcal{M}^n$ which may be cast locally into the form
\begin{equation}
g = \frac{dr^2}{r^4 f(r)^2} + \frac{1}{f(r)}{}^{(n)}h \label{warpprod}
\end{equation}
 where $h$ is an $n$-dimensional metric on $\mathcal{M}$. We will take local coordinates $x^\mu = ( r,x^i)$, with Greek indices ranging over $0, \ldots, n$ and Roman indices over $1, \ldots, n$ and we will use the notation ${}^{(n)}\Gamma_i{}^j{}_k, {}^{(n)}R^i{}_{jkl}, {}^{(n)}R_{ij}$ to refer to objects intrinsic to $\mathcal{M}$. We will at this stage avoid the assumption that $f$ is positive or that $h$ is Riemannian. We wish to calculate the projective tensor:
 \begin{equation}
W^\mu{}_{\nu \sigma \rho} = R^\mu{}_{\nu \sigma \rho} + \frac{1}{n}\left ( \delta^\mu{}_\rho R_{\nu \sigma} - \delta^\mu{}_\sigma R_{\nu \rho}\right). \label{projcurv}
\end{equation}
The non-zero components of the Riemann tensor are:
\begin{eqnarray}
R^i{}_{jkl} &=& {}^{(n)}R^i{}_{jkl} + \frac{r^4 f'^2}{4 f} \left(\delta^i{}_l h_{jk} - \delta^i{}_k h_{jl} \right) \\
R^0{}_{i0j} &=& \left( r^3 f' + \frac{1}{2} r^4 f'' - \frac{r^4 f'^2}{4f}\right) h_{ij} \label{riemcomp}
\end{eqnarray}
Contracting $R_{\mu \nu} = R^\tau{}_{\mu \tau \nu}$ to get the Ricci tensor, we find the non-zero components are:
\begin{eqnarray}
R_{00} &=& n \left(\frac{f''}{2 f} + \frac{f'}{f r} - \frac{f'^2}{4f^2}\right) \\
R_{ij} &=& {}^{(n)}R_{ij} + \left( r^3 f' + \frac{1}{2} r^4 f'' - \frac{n}{4}\frac{ r^4 f'^2}{f}\right).
\end{eqnarray}
We have finally the necessary information to calculate the projective curvature (\ref{projcurv}) and we find that the non-zero components are:
\begin{eqnarray}
W^i{}_{jkl} &=& {}^{(n)}R^i{}_{jkl}  + \frac{1}{n} \left( \delta^i{}_l {}^{(n)}R_{jk} - \delta^i{}_k {}^{(n)}R_{jl}\right) + \frac{1}{n} \left(r^3 f' + \frac{1}{2} r^4 f'' \right)\left(\delta^i{}_l h_{jk} - \delta^i{}_k h_{jl} \right) \\
W^0{}_{i0j} &=& \left(1-\frac{1}{n} \right)\left(r^3 f' + \frac{1}{2} r^4 f''\right) h_{ij} - \frac{1}{n} {}^{(n)}R_{ij}
\end{eqnarray}
\emph{The dependence on $f$ is solely through $f'$ and $f''$, so that adding a constant to $f$ does not change the projective tensor.}

\subsection{An application -- metrics of constant curvature}

Let us suppose that $h$ is a metric of constant curvature , so that we may write
\begin{equation}
{}^{(n)}R^i{}_{jkl} = K(\delta^i{}_k h_{j l} - \delta^i{}_l h_{jk}), \qquad {}^{(n)}R_{ij} = (n-1)Kh_{ij} 
\end{equation}
then the projective tensor simplifies to give:
\begin{eqnarray}
W^i{}_{jkl} &=& \frac{1}{n} \left(r^3 f' + \frac{1}{2} r^4 f''  - K\right)\left(\delta^i{}_l h_{jk} - \delta^i{}_k h_{jl} \right) \\
W^0{}_{i 0 j} &=& \left(1-\frac{1}{n}\right)\left(r^3 f' + \frac{1}{2} r^4 f''  - K\right)h_{ij}
\end{eqnarray}
We already know that the projective tensor vanishes if and only if the metric $g$ is constant curvature. In our case, this happens when $f$ satisfies the differential equation:
\begin{equation}
r^3 f' + \frac{1}{2} r^4 f''  = K.
\end{equation}
This linear equation is readily solved to give
\begin{equation}
f(r) = \frac{K}{r^2} - \frac{2 a}{r}+b
\end{equation}
with $a$ and $b$ arbitrary constants. Using the components of the Riemann tensor derived above (\ref{riemcomp}) we check and find that $g$ is indeed constant curvature, with
\begin{equation}
R^\mu{}_{\nu \sigma \tau} = \left(bK - a^2 \right) \left( \delta^\mu{}_\sigma g_{\nu \tau} - \delta^\mu{}_\tau g_{\nu \sigma}\right) \label{constcurv}
\end{equation}
so we can construct constant curvature metrics sliced by other constant curvature metrics, and can rule out the possibility of such constructions. Some caution must be exercised here, since changing the signature changes the sign of the right hand side of (\ref{constcurv}) but not the left. We will take the view that the sign of the curvature is determined by (\ref{constcurv}) when $g$ has mostly plus signature, since this means AdS has negative constant curvature and dS positive. For example, if we take $K=1, b=-1, a=0$ then $f$ is negative. In 3 dimensions, after a signature change, we have the metric
\begin{equation}
ds^2 = - \frac{dr^2}{(r^2-1)^2} + \frac{r^2}{r^2-1}\left(d\theta^2+\sin^2\theta d\phi^2 \right)
\end{equation}
which for $r>1$ is Lorentzian, with mostly plus signature. Since we changed signature, this is constant \emph{positive} curvature ($-bK=1$). Setting $r=\coth t $ we recover the standard FRLW $k=1$ metric for de Sitter
\begin{equation}
ds^2 = -dt^2 + \cosh^2 t \left(d\theta^2+\sin^2\theta d\phi^2 \right)
\end{equation}
It is straightforward using the results of this section to find all of the standard metrics for constant curvature spaces.

\section{Constants of the motion}

In \cite{Topalev} Topalov and Matveev showed using results of Painlev\'e and Levi-Civita that if two metrics, $g$ and $\overline{g}$,  on an $n$-dimensional manifold are projectively equivalent, then it is possible to construct $n$ integrals of the geodesic flow of each metric. In the case where the metrics are strictly non-proportional at some point $x$, i.e.\ $\det{(g-\lambda \overline{g})}$ has only simple roots at $x$ then they showed in addition that this implied that the geodesic flow is integrable for both metrics. We can explicitly construct these integrals in the case when the metrics may be cast into the form (\ref{warpprod}), with corresponding functions $f, \overline{f}$. We first briefly review the construction of \cite{Topalev} before stating the result.

Given $g, \overline{g}$ metrics on $M^{n+1}$, we form an endomorphism of $TM$,  $G = g^{-1} \overline{g}$. The characteristic polynomial $\det (G - \mu I) = c_0 \mu^{n+1} + c_1 \mu^n+\ldots+c_{n+1}$ has coefficients $c_i$ which are smooth functions on the manifold. From them we may construct $S_k = \sum_{i=0}^k c_i G^{k-i}$, for $k=0, \ldots, n$, which are endomorphisms of $TM$. Finally, the functions
\ben
I_k = \left( \det G\right)^{-\frac{k+2}{n+1}} \overline{g}(S_k \dot{x}, \dot{x})
\een
are conserved along a geodesic $x(t)$ of the metric $g$ and mutually commute.

If we now specialize to the case of metrics of the form (\ref{warpprod}), one may show after some calculation that the integrals $I_k$ may be reduced to the following form:
\ben
I_k = (-1)^{n+1} \left[2 (-1)^k \binom{n}{k} E_g + (\overline{f}-f) J_g^2 \sum_{i=0}^k (-1)^i \binom{n}{i} \right ] \label{Ik}
\een
where
\ben
E_g = \frac{1}{2} \left( \frac{\dot{r}^2}{r^4 f(r)} + \frac{h_{ij}\dot{x}^i \dot{x}^j}{f}\right) \qquad \mathrm{and} \qquad J_g^2 = \frac{h_{ij}\dot{x}^i \dot{x}^j}{f^2}
\een
are conserved quantities under the geodesic flow of $g$ and $\overline{f}-f$ must be a constant by the geodesic equivalence of the metrics $g, \overline{g}$. Thus we see that by this construction, we recover the conserved quantities of the geodesic flow, but do not gain any extra. 

In \cite{Topalev2} the authors of \cite{Topalev} showed that these classical constants of the motion have quantum mechanical analogues. This means that for each $I_k$ which may be written in the form:
\ben
I_k = I_k{}^{\mu \nu} p_\mu p_\nu, \label{stakkil}
\een
where $p_\mu$ are the canonical momenta of the geodesic flow, there exists an operator
\ben
\mathcal{I}_k = \nabla_\mu I_k{}^{\mu \nu} \nabla_\nu
\een
which commutes with the scalar Laplacian. It is not true in general that a classical constant of the motion of the form (\ref{stakkil}) arising from a St\"ackel-Killing tensor gives rise to such an operator. In general there is an anomaly, as found by Carter \cite{Carter:1977pq}. We may calculate the form of these operators for our family of metrics and we find that they take precisely the form (\ref{Ik}) with $E_g$ and $J_g^2$ replaced by the operators:
\ben
\mathcal{E}_g = \frac{1}{\sqrt{g}} \pd{}{x^\mu} \sqrt{g} g^{\mu \nu} \pd{}{x^\nu} \qquad \mathrm{and} \qquad \mathcal{J}_g^2 = \frac{1}{\sqrt{h}} \pd{}{x^i} \sqrt{h} h^{ij} \pd{}{x^j}
\een
respectively. Here $\mu, \nu$ range over $0, \ldots, n$, where $0$ refers to the (not necessarily timelike) $r$ direction and $i, j$ range over $1, \ldots n$. These clearly commute with the scalar Laplacian, which is in fact given by $\mathcal{E}_g$ and hence $\mathcal{I}_k$ give conserved quantum numbers.

\section{Conclusion}

We have shown that the fact the the light rays of the Kottler metric are independent of the cosmological constants follows from the projective equivalence of the optical metrics for all values of $\Lambda$. Motivated by this we have found a large collection of one parameter families of metrics which are projectively equivalent for any value of the parameter. As a sub-case we find the Kottler optical metrics together with the constant curvature metrics of any signature. Using results of Matveev and Topalov we showed that this implies the existence of classical and quantum conserved quantities and we related these to known quantities.

\end{document}